\documentclass[final,5p,times,twocolumn]{elsarticle}

\usepackage{epsfig}

\usepackage{amssymb}

\journal{Physica E}

\begin{document}

\begin{frontmatter}



\title{Theory of the electronic and transport properties of graphene under a periodic electric or magnetic field}


\author{Cheol-Hwan Park}
\author{Liang Zheng Tan}
\author{Steven G. Louie}
\ead{sglouie@berkeley.edu}
\address{Department of Physics, University of California at Berkeley,
Berkeley, California 94720}
\address{Materials Sciences Division, Lawrence Berkeley National
Laboratory, Berkeley, California 94720}

\begin{abstract}
We discuss the novel electronic properties of graphene
under an external periodic scalar or vector potential,
and the analytical and numerical methods used to investigate them.
When graphene is subjected to a one-dimensional periodic scalar potential,
owing to the linear dispersion and the chiral (pseudospin)
nature of the electronic states,
the group velocity of its carriers is renormalized highly anisotropically
in such a manner that the velocity is invariant
along the periodic direction but is
reduced the most along the perpendicular direction.
Under a periodic scalar potential, new massless Dirac fermions
are generated at the supercell Brillouin zone boundaries.
Also, we show that if the strength of the applied
scalar potential is sufficiently strong,
new zero-energy modes may be generated.
With the periodic scalar potential satisfying some special
conditions, the energy dispersion near the Dirac point
becomes quasi one-dimensional.
On the other hand, for graphene under a one-dimensional
periodic vector potential (resulting in a periodic
magnetic field perpendicular to the graphene plane),
the group velocity is reduced
isotropically and
monotonically with the strength of the potential.
\end{abstract}


\end{frontmatter}


\section{Introduction}
\label{sec:intro}

\begin{figure}
\begin{center}
\includegraphics[width=1.0\columnwidth]{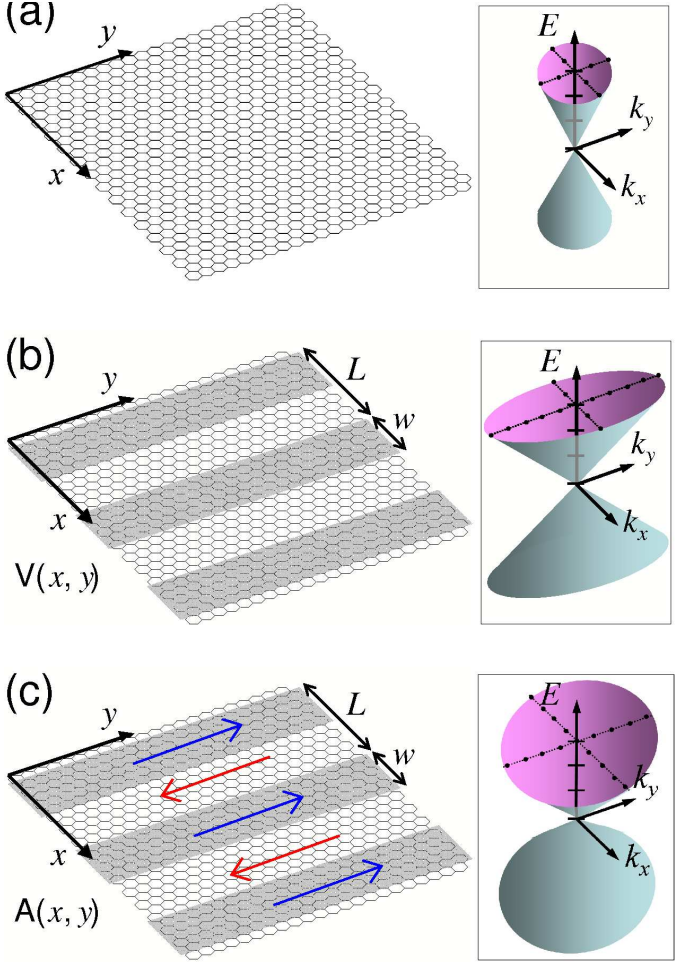}
\end{center}
\caption{
(a) Schematic diagram of graphene. Inset:
the linear and isotropic energy dispersion near one of the Dirac points
in graphene.
(b) A 1D graphene superlattice
formed by Kronig-Penney type of scalar potential $V(x,y)=V(x)$
periodic along the {\it x} direction with spatial period $L$.
The potential is $U_0/2$ in the grey regions and $-U_0/2$ outside.
Inset: energy dispersion of charge carriers in this graphene superlattice.
The energy dispersion along any line in 2D wavevector
space going through the Dirac point is linear but with different group velocity.
For a particle moving parallel to the
periodic direction, the group velocity is
not renormalized at all whereas that for a particle moving
perpendicular to the direction of periodicity is reduced the most.
(c) A 1D graphene superlattice formed by Kronig-Penney type of vector
potential ${\bf A}(x,y)=A_y(x)\,\hat{y}$
periodic along the {\it x} direction with spatial period $L$.
The vector potential $A_y(x)$ is $A_0/2$ in the grey regions
and $-A_0/2$ outside.
Inset: energy dispersion of charge carriers in this graphene superlattice.
The group velocity around the Dirac point is reduced {\it isotropically}.}
\label{Fig1}
\end{figure}

Graphene is a single atomic layer of carbon atoms arranged in a honeycomb structure.
The electronic states in graphene obey a unique linear energy dispersion relation
near the Fermi energy [Fig.~1(a)], and they possess an additional
quantum number called pseudospin which describes the electron's probability
amplitudes at the two different sublattices of carbon atoms forming
graphene~\cite{neto:109}.
These behaviors are similar to those of massless neutrinos in relativistic
quantum physics except that the role played by the actual spin of the neutrinos
is now replaced by the pseudospin in graphene.

When graphene is subjected to a slowly varying nanoscale external periodic scalar or vector
potential, the quasiparticles in graphene show even more interesting
physics.
Recently, there have been several studies on the electronic
properties of graphene under either a periodic scalar
potential~\cite{bai:2007PRB_Graphene_SL,park:2008NatPhys_GSL,barbier:115446,park:2008NL_Supercollimation,
park:126804,PhysRevB.79.115427,PhysRevLett.103.046808,PhysRevLett.103.046809,PhysRevB.81.075438,PhysRevB.81.205444},
under a periodic vector
potential~\cite{masir:235443,masir:035409,ghosh:arxiv,
dell'anna:045420,0953-8984-20-48-485210,PhysRevB.80.054303,tan:arxiv},
or under a periodic
corrugation~\cite{isacsson:035423,guinea:075422,wehling:EPL2008_graphene_corrugation}.

Graphene superlattices are not just theorists' dream but
have been experimentally
realized. Superlattice patterns with periodicity
as small as 5~nm
have been imprinted on graphene through electron-beam
induced deposition of adsorbates~\cite{meyer:123110},
triangular patterns with $\sim$10~nm lattice period
have been observed for graphene on metal
surfaces~\cite{marchini:2007PRB_Graphene_Ru,vazquez:2008PRL_Graphene_SL,
sutter:2008NatMat,martoccia:126102,
coraux:2008NL,ndiaye:2008NJP,pletikosic:056808},
and periodically
corrugated graphene sheet has also been reported~\cite{lau:ripple}.
Periodically patterned gates
can provide another route for making
graphene superlattices.

In this paper, we review the
electronic properties of charge carriers in graphene under an
external periodic scalar or vector potential.
Especially, we focus on one-dimensional (1D)
periodic potentials [e.\,g.\,, Figs.~1(b) or~1(c)] for simplicity.
However, many of the essential findings discussed here are applicable
to two-dimensional (2D) periodic potentials.
We also discuss the methodologies used in the analytical and numerical
calculations.

The rest of the paper is organized
as follows.  In Secs.~\ref{sec:scalar} and~\ref{sec:vector},
we present the analytical derivation of the energy-momentum
dispersion relation near the Dirac points (original or
newly generated) in graphene under an external periodic
scalar potential and vector potential, respectively.
In Sec.~\ref{sec:num_calc},
we present the details of our numerical calculations
used in studying the effects of stronger perturbing potentials.
In Sec.~\ref{sec:dis}, we discuss the emerging massless Dirac fermions
and quasi-1D modes under a strong external periodic scalar potential.
Finally, in Sec.~\ref{sec:summ}, we summarize our findings.

\section{Graphene under an external periodic scalar potential}
\label{sec:scalar}

In this section, through analytical calculations,
we show that when a 1D periodic scalar potential is applied
to graphene: (i) the group velocity of the massless Dirac fermions
is anisotropically renormalized in momentum space in an unexpected fashion,
and (ii) new massless Dirac fermions are
generated at the supercell Brillouin zone boundaries~\cite{park:126804}.

We consider a situation where the spatial variation of the external periodic
potential is much slower than the inter-carbon distance
so that inter-valley scattering between the {\bf K} and {\bf K}$'$
points in the Brillouin zone
may be neglected~\cite{ando:1998JPSJ_NT_Backscattering,mceuen:1999PRL_NT_Backscattering}.
We shall further limit our discussion to
the low-energy electronic states of graphene which have wavevector
${\bf k}+{\bf K}$ close to the {\bf K} point, i.\,e.\, with $|{\bf k}|\ll|{\bf K}|$.

There are two carbon atoms per unit cell in graphene,
forming two different sublattices. Hence the
eigenstate of charge carriers in graphene can
be represented by a two component basis  vector.
The Hamiltonian of the low-energy quasiparticles in pristine
graphene in a pseudospin basis,
$\left(\begin{array}{cc}1\\0\end{array}\right)e^{i{\bf k}\cdot{\bf r}}$
and $\left(\begin{array}{cc}0\\1\end{array}\right)e^{i{\bf k}\cdot{\bf r}}$,
where $\left(\begin{array}{cc}1\\0\end{array}\right)$ and
$\left(\begin{array}{cc}0\\1\end{array}\right)$
symbolically represent Bloch sums
of $\pi$-orbitals with wavevector {\bf K} on the sublattices A and B,
respectively,
is given by~\cite{wallace:1947PR_BandGraphite}
\begin{equation}
H_0=\hbar v_0\left(-i\sigma_x\partial_x-i\sigma_y\partial_y\right)
\,,
\label{eq:H_0}
\end{equation}
where $v_0$ is the band velocity
and the $\sigma$'s are the Pauli matrices.
The eigenstates and the energy eigenvalues are given by
\begin{equation}
\psi^0_{s,{\bf k}}({\bf r})=\frac{1}{\sqrt{2}}\left( \begin{array}{c}
1\\
se^{i\theta_{\bf k}}\end{array}\right)e^{i{\bf k}\cdot{\bf r}}
\label{eq:solH0}
\end{equation}
and
\begin{equation}
E^0_{s}({\bf k})=s\hbar v_0 k\,,
\label{eq:E}
\end{equation}
respectively, where $s=\pm1$ is the band index and $\theta_{\bf k}$ is
the angle between {\bf k} and the $+k_x$ direction.

We now assume that a 1D scalar potential $V(x)$,
periodic along the $x$ direction with periodicity $L$,
is applied to graphene [Fig.~1(b)].
The Hamiltonian $H$ then reads
\begin{equation}
H=\hbar v_0\left(
-i\sigma_x\partial_x-i\sigma_y \partial_y+
{I}\ V(x)/\hbar v_0\right),
\label{eq:H}
\end{equation}
where $I$ is the $2\times2$ identity matrix.
Next we perform a similarity transform, $H'=U_1^\dagger HU_1$,
using the unitary matrix
\begin{equation}
U_1=\frac{1}{\sqrt{2}}\left(\begin{array}{cc}
e^{-i\alpha(x)/2} & -e^{i\alpha(x)/2}\\
e^{-i\alpha(x)/2} & e^{i\alpha(x)/2}
\end{array}\right)
\label{eq:U_1}
\end{equation}
where $\alpha(x)$ is given by
\begin{equation}
\alpha(x) = 2\int_0^xV(x')\,dx'/\hbar v_0\,.
\label{eq:alpha}
\end{equation}
Here, without losing generality, we shall assume that an
appropriate constant has been
subtracted from $V(x)$ and that $V(x)$ has been shifted
along the {\it x} direction so that the averages of
both $V(x)$ and $\alpha(x)$ are zero.
The transformed Hamiltonian $H'$ takes the form
\begin{equation}
H'=\hbar v_0\left(\begin{array}{cc}
-i\partial_x & -e^{i\alpha(x)}\partial_y\\
e^{-i\alpha(x)}\partial_y & i\partial_x
\end{array}\right)\,.
\label{eq:H_prime}
\end{equation}
A similar transform has been used to study
nanotubes under a sinusoidal
potential~\cite{PhysRevLett.87.276802,novikov:235428}.

We are interested in the low-energy quasiparticle states
whose wavevector
${\bf k}\equiv{\bf p}+{\bf G}_m/2$ [where
${\bf G}_m=m\left({2\pi}/{L}\right)\hat{x}\equiv m\,G\, \hat{x}$
is a reciprocal vector]
is such that
$|{\bf p}|\ll G$. In this case,
we could treat the terms containing $\partial_y$ in Eq.~(\ref{eq:H_prime})
as a perturbation.
Also, to a good approximation, $H'$ may be reduced to a
$2\times2$ matrix using the following two states as basis functions
\begin{equation}
\left(\begin{array}{c}
1\\
0
\end{array}\right)'e^{i({\bf p}+{\bf G}_m/2)\cdot{\bf r}}\ {\rm and }\ \
\left(\begin{array}{c}
0\\
1
\end{array}\right)'e^{i({\bf p}-{\bf G}_m/2)\cdot{\bf r}}\,.
\label{eq:basis}
\end{equation}
Here, we should note that the spinors
$\left(\begin{array}{cc}1\\0\end{array}\right)'$ and
$\left(\begin{array}{cc}0\\1\end{array}\right)'$ now have
a different meaning from
$\left(\begin{array}{cc}1\\0\end{array}\right)$ and
$\left(\begin{array}{cc}0\\1\end{array}\right)$.

In order to calculate these matrix elements, we perform a
Fourier transform of $e^{i\alpha(x)}$
\begin{equation}
e^{i\alpha(x)}=
\sum^{\infty}_{l=-\infty}f_l[V]e^{\,i\,l\,G\,x},
\label{eq:e_alpha}
\end{equation}
where the Fourier components $f_l[V]$'s are determined by the
periodic potential $V(x)$.
We should note that in general
\begin{equation}
|f_l|<1\,,
\label{eq:f}
\end{equation}
which can directly be deduced from Eq.~(\ref{eq:e_alpha}).
The physics simplifies when the external potential $V(x)$ is an even
function and hence $\alpha(x)$ in Eq.~(\ref{eq:alpha}) is an
odd function.
If we take the complex conjugate of Eq.~(\ref{eq:e_alpha})
and change $x$ to $-x$,
it is evident that $f_l[V]$'s are real.
General cases other than even potentials
are discussed in Ref.~\cite{park:126804}.
For states with wavevector ${\bf k}$ very close to ${\bf G}_m/2$, the $2\times2$ matrix $M$
whose elements are calculated from the Hamiltonian $H'$ with
the basis given by Eq.~(\ref{eq:basis})
can be written as
\begin{equation}
M=\hbar v_0\left(p_x\sigma_z+f_m p_y\sigma_y\right)
+\hbar v_0\,mG/2\cdot I
\,.
\label{eq:M}
\end{equation}
After performing yet another similarity transform
$M'=U_2^\dagger MU_2$
with
\begin{equation}
U_2=\frac{1}{\sqrt{2}}\left(\begin{array}{cc}
1 & 1\\
-1 & 1
\end{array}\right)\,,
\label{eq:U_2}
\end{equation}
we obtain the final result:
\begin{equation}
M'=\hbar v_0\left(p_x\sigma_x+f_m\, p_y\sigma_y\right)
+\hbar v_0\,m\,G/2\cdot I
\,.
\label{eq:M_prime}
\end{equation}
The energy eigenvalue of the matrix $M'$ is given by
\begin{equation}
E_{s}({\bf p})=s\hbar v_0 \sqrt{p_x^2+|f_m|^2p_y^2}+\hbar v_0\, m\, G/2\,.
\label{eq:E_p}
\end{equation}
Equation~(\ref{eq:E_p}) holds in general and not only for
cases where the potential $V(x)$ is even~\cite{park:126804}.
The only difference of the energy spectrum in Eq.~(\ref{eq:E})
from that in Eq.~(\ref{eq:E_p}),
other than a constant energy term,
is that the group velocity
of quasiparticles moving along
the $y$ direction has been changed from $v_0$ to
$|f_m| v_0$.
Thus, the electronic states near ${\bf k}={\bf G}_m/2$ are also those of
massless Dirac fermions but having a group velocity
varying {\it anisotropically} depending on the propagation direction.
The group velocity along the {\it x} direction is unchanged independent of the potential.
Moreover, the group velocity along the $y$ direction
is {\it always lower} than $v_0$ [Eq.~(\ref{eq:f})] regardless of the form or magnitude of the
periodic potential~$V(x)$ as schematically depicted in Fig.~1(b).

We have thus shown that other than the original Dirac points,
new massless Dirac fermions are generated around the supercell Brillouin zone boundaries,
i.\,e.\,, the case with non-zero {\it m} values in Eq.~(\ref{eq:E_p}).
It has also been shown that these newly generated
massless Dirac points are the only available states in a certain
energy window if graphene is subjected to a 2D
repulsive periodic scalar potential having triangular symmetry~\cite{park:126804}.

One more thing to note is that in graphene under an external periodic
scalar potential, a generalized pseudospin vector can be defined
and used to describe the scattering properties between eigenstates;
and especially, back-scattering processes by
a slowly varying impurity potential
are suppressed as in pristine graphene~\cite{park:126804}.

\section{Graphene under a 1D external periodic vector potential}
\label{sec:vector}

Now we move on to the case where a 1D vector potential
${\bf A}(x,y)=A_y(x)\,\hat{y}$
is applied to graphene [Fig.~1(c)].
We show through a novel transformation relation
between scalar and vector potentials~\cite{tan:arxiv} that,
unlike the electrostatic case, the group velocity
of charge carriers in graphene under a 1D periodic vector potential
is renormalized {\it isotropically}.

The superlattice Hamiltonian,
following the Peierls substitution, is given by
\begin{equation}
H=\hbar v_0\left(
-i\sigma_x\partial_x-i\sigma_y \partial_y
-\sigma_y eA_y(x)/\hbar c\right),
\label{eq:H_A}
\end{equation}
where {\it e} is the charge of an electron ($e<0$)
and {\it c} is the speed of light.
The time-dependent Dirac equation then reads
\begin{equation}
i\hbar\frac{d\psi}{dt}=
\hbar v_0\left(
-i\sigma_x\partial_x-i\sigma_y \partial_y
-\sigma_y eA_y(x)/\hbar c\right)\psi\,.
\label{eq:Dirac}
\end{equation}
Writing the wavefunction as
$\psi(x,y;t)=e^{-iEt/\hbar}\,e^{i\,k_y\,y}\,\varphi(x)$,
the Dirac equation becomes
\begin{equation}
E\varphi(x)=
\hbar v_0\left(
-i\sigma_x\partial_x+\sigma_y k_y
-\sigma_y eA_y(x)/\hbar c\right)\varphi(x)\,.
\label{eq:Dirac2}
\end{equation}
Now, if we multiply Eq.~(\ref{eq:Dirac2}) by $\sigma_y$ on both sides,
define $\varphi'(x)=U_3\,\varphi(x)$
with
\begin{equation}
U_3=\frac{1}{\sqrt{2}}\left(
\begin{array}{cc}
1 & 1\\
1 & -1\\
\end{array}
\right)\,,
\label{eq:U_3}
\end{equation}
and use the relations $U_3=U_3^\dagger=U_3^{-1}$,
$U_3\sigma_yU_3=-\sigma_y$, and $U_3\sigma_xU_3=\sigma_z$,
Eq.~(\ref{eq:Dirac2}) becomes
\begin{equation}
E'\varphi'(x)=
\hbar v_0\left(
-i\sigma_x\partial_x+\sigma_y\,k'_y
+I\,V'(x)/\hbar v_0\right)\varphi'(x)\,.
\label{eq:Dirac3}
\end{equation}
Here, we have defined
\begin{equation}
E'=-i\hbar v_0 k_y\,,
\label{eq:Eprime}
\end{equation}
\begin{equation}
k_y'=iE/\hbar v_0\,,
\label{eq:kyprime}
\end{equation}
and
\begin{equation}
V'(x)=-i\,e\,(v_0/c)\, A_y(x)\,.
\label{eq:Vprime}
\end{equation}
Equation~(\ref{eq:Dirac3}) is thus equivalent to
the Dirac equation with a periodic scalar potential
in Eq.~(\ref{eq:H}) except that now the variables
are imaginary numbers.  Using analytic
continuation~\cite{tan:arxiv}, we obtain the energy-momentum
dispersion relation in magnetic graphene superlattices
from that in electrostatic graphene superlattices.
For the states near the original Dirac point,
as shown in the previous section,
the energy dispersion in graphene under a periodic
scalar potential is given [from Eq.~(\ref{eq:E_p})]
by
\begin{equation}
E_{s}({\bf k})=s\hbar v_0 \sqrt{k_x^2+|f_0|^2k_y^2}\,,
\label{eq:E_p_m0}
\end{equation}
where, according to Eq.~(\ref{eq:e_alpha}),
\begin{equation}
f_0=\frac{1}{L}\int_0^L \exp\left(i\int_0^x\frac{2}{\hbar v_0}V(x')\,dx'\right)dx\,.
\label{eq:f0}
\end{equation}
Plugging Eqs.~(\ref{eq:Eprime}), (\ref{eq:kyprime}),
and~(\ref{eq:Vprime}) into Eq.~(\ref{eq:E_p_m0}), we obtain
\begin{equation}
E_s({\bf k})=s\frac{1}{|f'_0|}\,\hbar v_0\sqrt{k_x^2+k_y^2}\,,
\label{eq:E_MGS}
\end{equation}
where
\begin{equation}
f'_0 = \frac{1}{L}\int_0^L \exp\left(\int_0^x\frac{2e}{\hbar c}A_y(x')\,dx'\right)dx\,.
\label{eq:fprime}
\end{equation}
Therefore, from Eq.~(\ref{eq:E_MGS}), we find that
the group velocity in graphene under an external
periodic vector potential
(corresponding to a perpendicular magnetic field)
is renormalized {\it isotropically}
in the $k_x-k_y$ space~\cite{PhysRevB.80.054303,tan:arxiv}
even though the external vector potential
profile is highly anisotropic in the {\it x-y} plane [Fig.~1(c)].
Note from Eq.~(\ref{eq:fprime}) that
\begin{equation}
|f'_0|>1
\label{eq:fprime_1}
\end{equation}
regardless of the form of the
vector potential $A_y(x)$, i.\,e.\,,
the group velocity in graphene under an external
periodic vector potential is always reduced.
This result can in fact be used as a special case
to understand the predictions of velocity
reduction in metallic carbon nanotubes and gap reduction
in semiconducting carbon nanotubes
under a magnetic field~\cite{ajiki_ando,PhysRevB.68.155402}.

\section{Numerical calculation}
\label{sec:num_calc}
If one wants to find the energy eigenvalues and
eigenfunctions of graphene under an external periodic
potential with wavevector {\bf k} not very close to the
supercell Brillouin zone boundary centers
(${\bf G}_m/2$), one has to resort to
numerical calculations.  Such numerical calculations
have led us to the discovery of solutions
corresponding to new branches of massless
Dirac fermions~\cite{PhysRevLett.103.046808}
that are not found in the analytical
calculations discussed in the previous sections.

The scattering amplitudes arising from the periodic
scalar and vector potentials between eigenstates
of pristine graphene [Eq.~(\ref{eq:solH0})],
using the Hamiltonians in Eq.~(\ref{eq:H})
and in Eq.~(\ref{eq:H_A}), are given by
\begin{eqnarray}
&\left<\psi^0_{s,{\bf k}}\left|\, I\, V({\bf r})\,
\right|\psi^0_{s',{\bf k}'}\right>\nonumber\\
&=\sum_{{\bf G}}\frac{1}{2}
\left(1+ss'e^{i(\theta_{{\bf k}'}-\theta_{\bf k})}\right)V({\bf G})\ 
\delta_{{\bf k}',{\bf k}-{\bf G}}
\label{eq:scatter_V}
\end{eqnarray}
and
\begin{eqnarray}
&\left<\psi^0_{s,{\bf k}}\left|\,-ev_0/c\cdot
\sigma_y\, A_y({\bf r})\,\right|\psi^0_{s',{\bf k}'}\right>
\nonumber\\
&={ev_0}/{c}\cdot
\sum_{{\bf G}}\frac{i}{2}
\left(s'e^{i\theta_{{\bf k}'}}-se^{-i\theta_{{\bf k}}}\right)A_y({\bf G})\ 
\delta_{{\bf k}',{\bf k}-{\bf G}}\ ,
\label{eq:scatter_A}
\end{eqnarray}
respectively.
Here, ${\bf G}$ is a superlattice reciprocal lattice vector and
$V({\bf G})$ and $A_y({\bf G})$ are the corresponding
Fourier components of the external periodic scalar and vector
potentials, respectively.
Therefore, the energy dispersion and eigenstates of the
quasiparticles in a graphene superlattice are obtained non-perturbatively
within the single-particle picture
by solving the following set of linear equations
\begin{eqnarray}
&(E-E^0_{s,{\bf k}})\ c(s,{\bf k})\nonumber\\
&= \sum_{s',{\bf G}}
\frac{1}{2}\left(1+ss'e^{i(\theta_{{\bf k}'}-\theta_{\bf k})}\right)
V({\bf G})\ c(s',{\bf k}-{\bf G})
\label{eq:linear_V}
\end{eqnarray}
for graphene under an external periodic scalar potential,
and by solving the following set of linear equations
\begin{eqnarray}
&(E-E^0_{s,{\bf k}})\ c(s,{\bf k})\nonumber\\
&= \sum_{s',{\bf G}}
\frac{1}{2}\left(s'e^{i\theta_{{\bf k}'}}-se^{-i\theta_{{\bf k}}}\right)
A_y({\bf G})\ c(s',{\bf k}-{\bf G})\ ,
\label{eq:linear_A}
\end{eqnarray}
for graphene under an external periodic vector potential,
where $E$ is the superlattice energy eigenvalue
and we have used Eqs.~(\ref{eq:E}), (\ref{eq:scatter_V}), and~(\ref{eq:scatter_A}).
The amplitudes $c(s,{\bf k})$ and $c(s',{\bf k}-{\bf G})$
indicate the mixing among different unperturbed quasiparticle states
of pristine graphene.

Note that the scattering amplitude
methods presented here are
applicable to 2D graphene
superlattices in general and not just to 1D periodic systems.

\section{Emerging new massless Dirac fermions in a strong external periodic scalar potential}
\label{sec:dis}

\begin{figure*}
\begin{center}
\includegraphics[width=1.6\columnwidth]{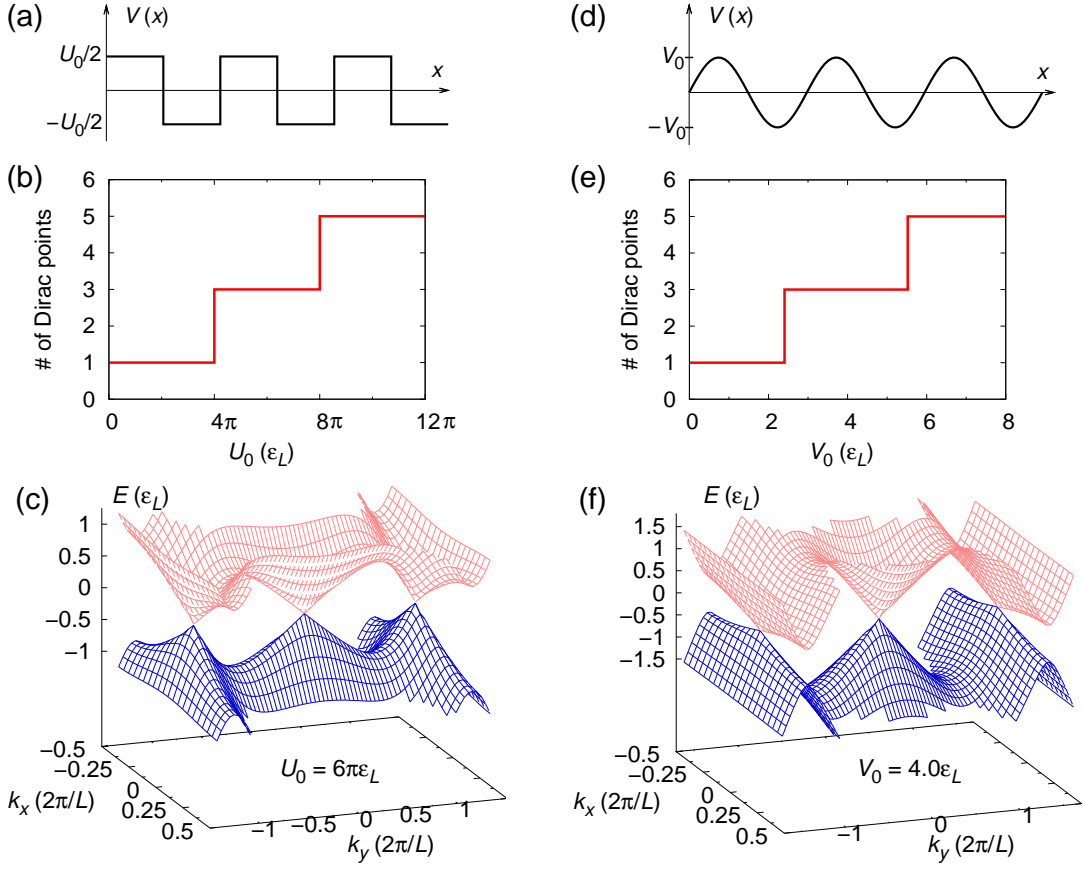}
\end{center}
\caption{(a) Schematic diagram of a Kronig-Penney type of scalar potential
applied to graphene given by $U_0/2$ for $0<x<L/2$
and $-U_0/2$ for $L/2<x<L$ with lattice period {\it L}.
(b) Number of Dirac points (not including spin and valley degeneracies)
in a graphene superlattice versus $U_0$.
(c) Electron energy in units of $\varepsilon_L$
($\equiv\hbar v_0/L$; for example, if $L=20$~nm, $\varepsilon_L=33$~meV)
versus wavevector near the Dirac point for a graphene superlattice
formed by the periodic scalar potential depicted in (a)
with $U_0=6\pi\varepsilon_L$.
(d)-(f): Same quantities as in (a)-(c) for a graphene superlattice
formed by a sinusoidal scalar potential $V(x)=V_0\sin(2\pi\,x/L)$.
In (f), $V_0=4.0\varepsilon_L$ was used.}
\label{Fig2}
\end{figure*}

If the external periodic scalar potential applied to graphene
is sufficiently strong, new branches of massless Dirac fermions are generated near the original Dirac
cone (i.\,e.\,, zero-energy
modes)~\cite{PhysRevLett.103.046808,PhysRevLett.103.046809,PhysRevB.81.075438,PhysRevB.81.205444}.
Note that these new zero-energy modes
are different from the
new massless Dirac fermions
discussed in Sec.~\ref{sec:scalar}, which have higher energies and
are generated at the supercell Brillouin zone
boundary centers (${\bf k}={\bf G}_m/2$)
no matter how weak the perturbing potential is.
As shown in Fig.~2, the number of zero-energy
Dirac modes increases
with the amplitude of the external periodic scalar potential.
These new zero-energy modes could have distinguishable signatures in
quantum Hall~\cite{PhysRevLett.103.046808} or transport measurement~\cite{PhysRevLett.103.046809}.

\begin{figure*}
\begin{center}
\includegraphics[width=2.0\columnwidth]{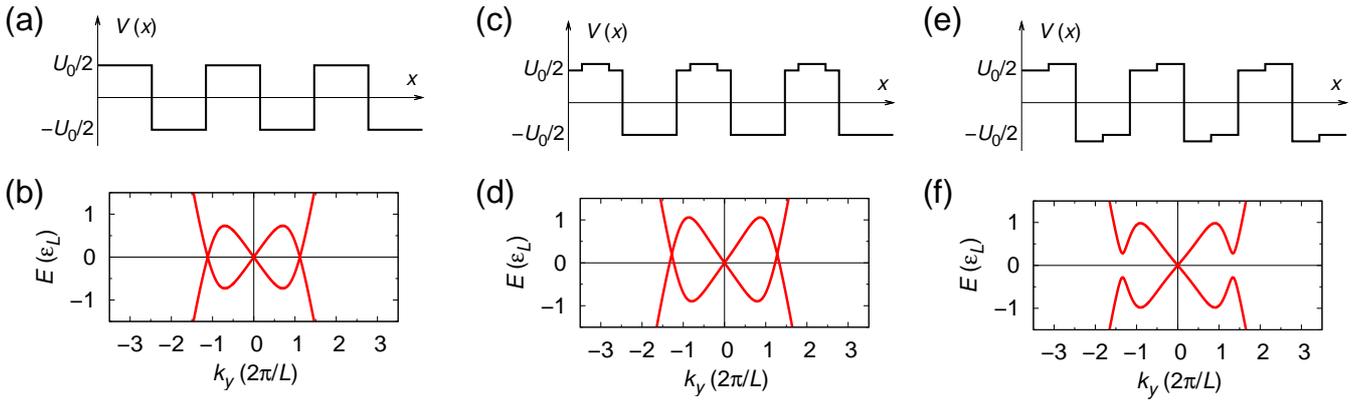}
\end{center}
\caption{(a) Kronig-Penney type of potential $V(x)$
given by $U_0/2$ for $0<x<L/2$ and $-U_0/2$ for $L/2<x<L$ with lattice period $L$.
(b) Electron energy (in units of $\varepsilon_L=\hbar v_0/L$)
versus $k_y$ with $k_x=0$ in a graphene superlattice formed by the periodic potential in (a)
with $U_0=6\pi\varepsilon_L$.
(c) and (d): Same quantities as in (a) and (b) for a periodic potential $V(x)$
with a perturbation that breaks the odd symmetry. The perturbing potential
$\Delta V(x)$ within one unit cell is given by $+10~\%$ of the potential
amplitude ($U_0/2$) for $L/8<x<3L/8$
and zero otherwise.
(e) and (f): Same quantities as in (a) and (b) for a periodic potential $V(x)$
with a perturbation that breaks the even symmetry. The perturbing potential
$\Delta V(x)$ within one unit cell is given by $+10~\%$ and $-10~\%$ of the potential
amplitude ($U_0/2$) for $L/4<x<L/2$ and for $L/2<x<3L/4$, respectively,
and zero otherwise.}
\label{Fig3}
\end{figure*}

The new zero-energy modes are generated when the applied
periodic scalar potential $V(x)$ has both even and odd
symmetries [Fig.~2 or Figs.~3(a) and~3(b)].
When the odd symmetry is broken [Fig.~3(c)],
new massless Dirac fermions are generated but the energy
position could be different from zero [Fig.~3(d)].
But, if the even symmetry is broken [Fig.~3(e)],
new massless Dirac fermions are not generated [Fig.~3(f)].
However, even in these broken-symmetry cases, the signatures of the
modified electronic bandstructure could be captured
in, e.\,g.\,, Landau level measurements~\cite{PhysRevLett.103.046808}.

\begin{figure*}
\begin{center}
\includegraphics[width=1.6\columnwidth]{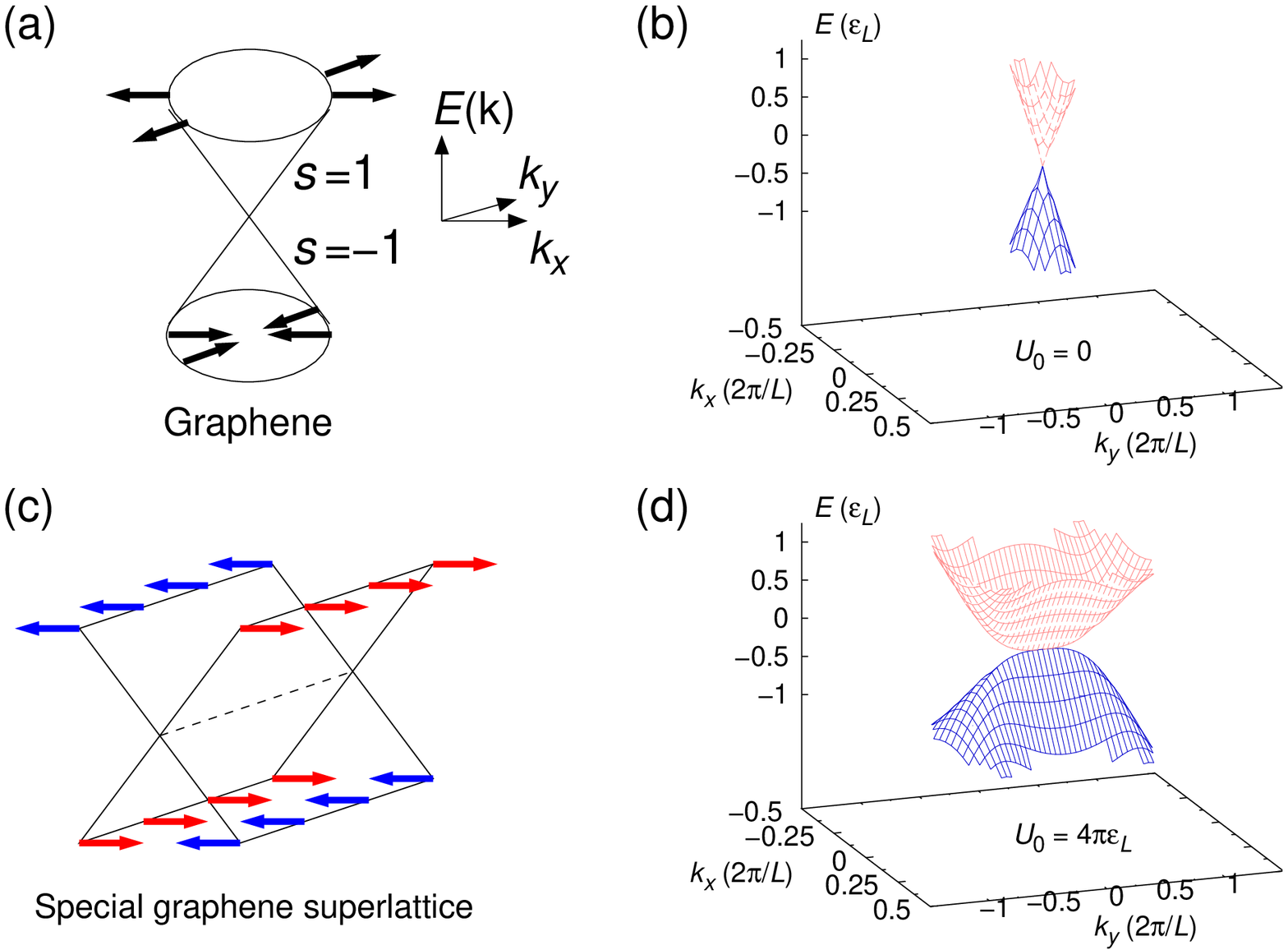}
\end{center}
\caption{
(a) Schematic diagram showing the electronic energy dispersion relations
and pseudospin vectors (black arrows) in graphene.
(b) Calculated electron energy bandstructure in graphene.
(c) and (d): Same quantity as in (a) and (b), respectively, for the considered
Kronig-Penney type of an SGS [as defined in Fig.~1(b) with $U_0=4\pi\varepsilon_L$
where $\varepsilon_L=\hbar v_0/L$].
Red and blue arrows in (c) represent the `right' and the `left'
pseudospin state, respectively.}
\label{Fig4}
\end{figure*}

It is worthwhile to focus on the conditions on the external
periodic potential under which the number of zero-modes jumps,
as shown by the steps in Fig.~2(b) and Fig.~2(e).
We call these systems at the jump special graphene superlattices
(SGSs)~\cite{park:2008NL_Supercollimation,PhysRevLett.103.046808}.
In an SGS, the group velocity along the $k_y$ direction vanishes.
Figure~4 shows that the energy dispersion in an SGS is
quasi-1D and the pseudospin in an SGS is either parallel or
antiparallel to the $+k_x$ (or periodic) direction~\cite{park:2008NL_Supercollimation}.
Because of this quasi-1D electron energy bandstructure in an SGS,
the group velocity of electronic states are almost the
same over a wide region in momentum space;
and this has led to the prediction that SGSs can be used for electron
beam supercollimation~\cite{park:2008NL_Supercollimation}.

\section{Summary}
\label{sec:summ}
The electronic structure of graphene
under a general external periodic scalar potential is modified from
that of graphene in several highly unexpected ways: (i) the group velocity
is anisotropically renormalized in momentum space
and (ii) new massless Dirac fermions are generated
at the supercell Brillouin zone boundary.
Moreover, when a strong 1D periodic scalar potential
is applied, new zero-energy modes emerge.
Under certain conditions, the group velocity of
charge carriers along the direction perpendicular to
the 1D periodic direction of the superlattice potential
vanishes.  In this special class of 1D graphene superlattices,
the electron energy bandstructure and pseudospin
structure are quasi-1D and these properties can
be used in collimating the electron flow.
With a 1D external periodic vector potential applied
to graphene, on the other hand, the group velocity
of charge carriers near the original Dirac point is
reduced isotropically.  The analytical and numerical
methods discussed in this paper can be used in
further investigating the novel properties of quasiparticles
in 1D and 2D graphene superlattices.


\section*{Acknowledgment}
We acknowledge Li Yang, Young-Woo Son, and Marvin L. Cohen
for fruitful discussions and collaboration.
The theoretical part of this work
was supported by NSF Grant No. DMR07-05941, and
L.Z.T. and the simulation part of the study
by the Director, Office of Science, Office of Basic Energy
Sciences, Division of Materials Sciences and Engineering Division,
U.S. Department of Energy under Contract No.\,DE-AC02-05CH11231.
C.-H.P. was supported by Office of Naval Research MURI Grant
No.\,N00014-09-1066.
Computational resources have been provided by TeraGrid and NERSC.











\end{document}